\documentclass[aps,prb,reprint,onecolumn,groupedaddress,showpacs]{revtex4-1}
\usepackage{graphicx}
\usepackage{amsmath}

\begin{document}
\title{Statistics of continuous weak quantum measurement of an arbitrary quantum system with multiple detectors}

\author{A. Franquet}
\email[]{A.FranquetGonzalez@tudelft.nl}
\affiliation{Kavli Institute of Nanoscience, Delft University of Technology, 2628 CJ Delft, The Netherlands}
\author{Hongduo Wei}
\affiliation{Department of physics, School of physics, Northwest University, Xi'An, P. R. China}
\author{Yuli V. Nazarov}
\affiliation{Kavli Institute of Nanoscience, Delft University of Technology, 2628 CJ Delft, The Netherlands}

\begin{abstract}In this paper, we establish a general theoretical framework for the description of continuous quantum measurements and the statistics of the results of such measurements. The framework concerns the measurement of an arbitrary quantum system with arbitrary number of detectors under realistic assumption of instant detector reactions and white noise sources. We attend various approaches to the problem showing their equivalence. The approaches include the full counting statistics (FCS) evolution equation a for pseudo-density matrix, the drift-diffusion equation for a density matrix in the space of integrated outputs, and discrete stochastic updates. We provide the derivation of the underlying equations from a microscopic approach based on full counting statistics method, a phenomenological approach based on Lindblad construction, and interaction with auxiliary quantum systems representing the detectors. We establish the necessary conditions on the phenomenological susceptibilities and noises that guarantee the unambiguous interpretation of the measurement results and the positivity of the density matrix. Our results can be easily extended to describe various quantum feedback schemes where the manipulation decision is based on the values of detector outputs. \end{abstract}

\maketitle

\section{Introduction}
The concept of quantum measurement is essential for the understanding and interpretation of quantum mechanics, and continuously inspires both theoretical and experimental research \cite{reviewnoise}. The proper description of the measurement process and setup is essential in  the quantum realm. The projective measurement \cite{Neumann}, although can be realized experimentally, is not the only way to acquire information about the state of a quantum system. One of the most experimentally relevant situations is the setup and paradigm of {\it continuous weak linear measurement} (CWLM) \cite{CWLM0,CWLM1,CWLM2,Wei,CWLM25,CWLM26,CWLM4}. 
In this setup, a weak coupling between the quantum system and its environment results in continuous entanglement of the system and the environmental degrees of freedom, those include the detector variables. Thereby, the (discrete) quantum information from the measured system is converted to continuous time-dependent detector outputs. At the same time, the environment induces the decoherence and relaxation of the quantum system, which is an inevitable feedback of the measurement process.  The measurement results are random incorporating intrinsic noises of the detectors, and their statistics is interesting and important to reveal quantum features of the system measured. Recent experimental advances enable faster and more accurate CWLM and even permit combination of CWLM and projective measurement. \cite{Devoret, Huard, DiCarlo, SiddiqiSingle, SiddiqiEntanglement}. This allows to experimentally access the statistics in question and makes it relevant to describe and predict the statistics for arbitrary complex CWLM setups.

There are various approaches to statistics of CWLM. In Ref. \cite{Wei} an approach based on FCS has been developed and applied for several simple situations, in particular, the qubit purification has been demonstrated. Recently, the same approach has been extended to describe the situation of conditioned measurement where a CWLM ends with a projective measurent. This has been done for a single \cite{Albert1} and two \cite{Albert2} detectors and connection with the theory of weak values \cite{weakvalues} has been established. Many authors prefer the so-called Bayesian approach to the description of quantum measurement where one implements a stochastic update of density matrix \cite{trajectory,CWLM1,SiddiqiSingle} that does not immediately provide a closed expression for the statistics but permits rather efficient numerical simulations. Recent advances in this direction are presented in \cite{Atalaya2018,Dressel2017}. There is still no general scheme unifying the approaches, neither the equivalence between approaches has been shown generally and explicitly. For instance, Ref. \cite{Ruskov2010}  basically repeats the numerical calculations of Ref. \cite{Wei} with a different method. The generalization of the descriptions on an arbitrary number of detectors and arbitrary complex quantum system has not been done yet.

The goal of this article is to establish a general framework for the description of the CWLM in the case of an arbitrary number of detectors and arbitrary quantum system measured. The only important restriction on the applicability of the framework is the assumption that the time correlation of noises and time delays of the susceptibilities take place at a smaller scale than the typical scale of quantum evolution. This results in simple Markovian evolution equations and update schemes. This is also a usual experimental situation.

In the article, we consider three alternative descriptions of the statistics of the measured results employing three different derivation methods and showing their equivalence. First description gives the generating function of the statistics in terms of a solution of an evolution equation for a pseudo-density matrix, such equations are common in FCS context \cite{QuantumTransport} 
Second description is a drift-diffusion equation for a density matrix in the space of integrated detector outputs. Third description involves a stochastic update of the density matrix and summation over random trajectories in the space of integrated outputs. We derive these results with a microscopic method based on FCS approach, а phenomenological method that employs Lindblad construction \cite{Lindblad1976}, and, in the context of the update, a method where the detection is modeled with axillary quantum systems. We establish conditions on noises and susceptibilities involved that i. guarantee the unambiguous interpretation of the detector outputs ii. guarantee the positivity of the density matrix. We specify the minimum detection feedback on the quantum system measured.

This does not exhaust all possible approaches and formulations. Various path integral methods \cite{Mensky1994, Wei, Chantasri2015} are beyond the scope of this article. The potential importance of these methods is their ability to capture the physics beyond Markov approximation, and we believe they are redundant for Markovian setups. We note that the methods described in the article allow for simple non-Markovian extensions in case of delay in classical variables. Similar extensions are plausible for the description of quantum feedback schemes where the feedback does depend on the accumulated value of detector outputs. This will be discussed in detail in future publications. 

The paper is organized as follows. In Section \ref{sec:FCS} we provide the full microscopic derivation of the multi-detector measurement and demonstrate that the measurement statistics are completely described by an evolution equation for a pseudo-density matrix. We establish the necessary conditions for the unambiguous interpretation of the measurement results and for the positivity of the density matrix. In Section \ref{sec:driftdiffusion}, we show the equivalence of this scheme and a drift-diffusion equation for a density matrix that encompasses the integrated detector outputs.  In Section \ref{sec:lindblad}, we reverse-engineer the drift-diffusion equation providing its phenomenological derivation. Thereby we establish the minimum detector feedback on the measured system.
At this stage, it is convenient to rescale the outputs and separate the measuring part of the system into independent detectors. This is achieved by a linear transformation diagonalizing the matrix of the detector noises (Section \ref{sec:separation}). In Section \ref{sec:discreteupdate}, we turn to another approach introducing a general stochastic discrete update process that is equivalent to the drift-diffusion equation. In Section \ref{sec:trajectories} we demonstrate that the process 
is equivalent to the averaging over stochastic trajectories in the space of the integrated outputs with the trajectory-conditioned density matrix of the system measured. In two subsections of this Section, we specify two concrete realizations of the stochastic update: oscillator and qubit update. We conclude in Section \ref{sec:conclusions}

\section{FCS derivation}
\label{sec:FCS}
In this Section, we will derive an equation that determines statististics of time-integrated outputs of a set of detector variables $\hat{V}_i(t)$, Latin index $i$ numbering the detectors. We will follow the approach of \cite{Kindermann} in the description of the measurement and extend it to the case of multiple detectors. The key element of the approach is to introduce a pair of extra canonically conjugated variables $\hat{\chi}_i,\hat{s}_i$ for each detector. Their operators satisfy the canonical commutation relations $[\hat{\chi}_i,\hat{s}_j]=i\delta_{ij}$, $\hat{s}_i,\hat{\chi}_i$ being analogous to the momentums and coordinates, respectively.  The coupling Hamiltonian of these extra variables and the detector variables is postulated to be $H_c = - \hat{\chi}_i \hat{V}_i$ (we assume summation over repeating indices) and there are no other Hamiltonian terms involving $\hat{\chi}_i,\hat{s}_i$. This guarantees that the operators $\hat{s}$ represent an integrated detector output, since by virtue of Heisenberg equations
\begin{equation}
\frac{d \hat{s}_i}{dt} = \hat{V}_i(t).
\end{equation}
To proceed, let us consider the evolution of the density matrix of the detectors in variables $\boldsymbol{\chi} \equiv \{\chi_i\}$ in a time interval $(t_1,t_2)$. Such representation is especially convenient since $d\hat{\chi}_i/dt =0$ so that these variables do not change upon the evolution. Following the lines of \cite{Kindermann}, we obtain the relation between initial and final density matrices of the detectors($\hat{R}$ here is the initial density matrix of the whole system)
\begin{equation}
\rho_f(\boldsymbol{\chi}^+,\boldsymbol{\chi}^-) = P(\boldsymbol{\chi}^+,\boldsymbol{\chi}^-) \rho_{in}(\boldsymbol{\chi}^+,\boldsymbol{\chi}^-).
\end{equation}
 The matrices are thus related by so-called FCS kernel $P(\boldsymbol{\chi}^+,\boldsymbol{\chi}^-)$ that is given by 
 \begin{equation}
 \label{eq:FCSkernel}
 P(\boldsymbol{\chi}^+,\boldsymbol{\chi}^-) = \mathop{{\rm Tr}}\limits_{\rm sys}
  \overrightarrow{T} \exp\{-{i} \int_{t_1}^{t_2}{dt \bigl[ \hat{H}_{\rm sys}-\chi^+_i \hat{V}_i 
  \bigr] }\}
   \hat{R}  \overleftarrow{T}\exp\{\;{i} \int_{t_1}^{t_2}{dt \bigl[ \hat{H}_{\rm sys}-\chi^-_i\hat{V}_i \bigr]}\} 
 \end{equation}
and $\overrightarrow{T}(\overleftarrow{T})$ denotes(inverse) time ordering. 

As explained in \cite{Kindermann}, if the Wigner representation of the density matrix,
\begin{equation}
 \rho(\boldsymbol{\chi},\boldsymbol{s}) = \int{\frac{d \boldsymbol{\zeta}}{2 \pi} \; e^{i\boldsymbol{s}\cdot \boldsymbol{\zeta}}\; \rho(\boldsymbol{\chi}+\frac{\boldsymbol{\zeta}}{2},\boldsymbol{\chi}-\frac{\boldsymbol{\zeta}}{2})},
\end{equation}
can be interpreted as a classical probability distribution $\Pi(\boldsymbol{\chi},\boldsymbol{s})$ for the detectors to be at a certain position $\boldsymbol{\chi}$ with momentum $\boldsymbol{s}$, the Wigner representation of the FCS kernel $P(\boldsymbol{\chi},\boldsymbol{s})$ can be interpreted as the probability distribution of
the shifts in momentum $\boldsymbol{s}$, that is, as the distribution of integrated detector outputs  $\int_{t_1}^{t_2}\hat{V}_i(t) dt$.
This does not hold in general. Generally, a  Wigner representation cannot be interpreted as a probability distribution, so  the same applies to $P(\boldsymbol{\chi},\boldsymbol{s})$. In particular, $P(\boldsymbol{\chi},\boldsymbol{s})$ does not have to be positive. 

There is, however, an important case when the interpretation of the FCS kernel as the probability distribution of integrated detector outputs  is indeed applicable. In this particular case, $P(\boldsymbol{\chi},\boldsymbol{s})$ does not depend on $\boldsymbol{\chi}$. This implies  that $P(\boldsymbol{\chi}^+,\boldsymbol{\chi}^-)$ is a function of the difference of counting fields only,
$P(\boldsymbol{\chi}^+,\boldsymbol{\chi}^-) \equiv P(\boldsymbol{\chi}^+-\boldsymbol{\chi}^-)$. The latter function becomes the generating function of the probability distribution of the detector outputs.

In the following, we specify the model, compute the FCS kernel and reveal the conditions under which it depends on the difference of counting fields only. We argue that these conditions are met for any realistic measurement situation and therefore the FCS can be used for evaluation of the statistics of the integrated detector outputs.

We separate the whole system into a system to be measured and an environment. The system to be measured is a purely quantum system with finite number of degrees of freedom. We measure a set of operators $\hat{{\cal O}}_\alpha$ in the space of these degrees of freedom labeling them with Greek indices. They are coupled to the environmental degrees of freedom $\hat{Q}_\alpha$, the operators of the corresponding generalized forces,
\begin{equation}
H_c = - \hat{Q}_\alpha \hat{{\cal O}}_\alpha
\end{equation}
We will assume that in the absence of coupling the expectation values of the operators $\hat{V}_i, \hat{Q}_\alpha$ are absent, $\langle \hat{V}_i\rangle =0$, $\langle\hat{Q}_\alpha\rangle=0$ (if it is not so, we can always redefine the operators adding the constant terms compensating the averages).
If the coupling is sufficiently weak, the environment can be regarded as a linear one. The environment provides a reaction proportional to the first power of the operators  ${\cal O}_\alpha$. The detector variables $\hat{V}_i$ are also  defined as operators in the space of environmental degrees of freedom.
The total Hamiltonian thus reads:
\begin{equation}
H_{{\rm sys}} = H_{{\rm env}} +H_{{\rm q}} + H_{{\rm c}}
\end{equation}
where $H_{{\rm env}}$ and $H_{{\rm q}}$ define the dynamics of the environment and the system to be measured, respectively, and are operators in corresponding spaces. We employ this Hamiltonian to evaluate the FCS kernel (\ref{eq:FCSkernel}).

The answer would involve the correlators of the time-dependent operators $\hat{V}_i, \hat{Q}_\alpha$. It is instructive to assume that the correlations vanish at a time scale $t_c$ characterizing the environment while the quantum correlations in the system to be measured may persist at much larger scale. Let us separate the time interval $(t_1,t_2)$ into smaller intervals of duration ${\cal T} \gg t_c$. The dynamics of environment are independent for different intervals, so that the environmental degrees of freedom can be traced out separately within each interval. The duration ${\cal T}$ can be chosen such that the change of the density matrix of the system is small. 
Tracing out the environmental degrees of freedom in (\ref{eq:FCSkernel}) till time $t$,  results in a pseudo-density matrix $\tilde{\rho}(t)$ in the space of the system to be measured,
\begin{equation}
\tilde{\rho}(t) = \mathop{{\rm Tr}}\limits_{\rm env}
  \overrightarrow{T} \exp\{-{i} \int_{t_1}^{t}{d\tau\bigl[ \hat{H}_{\rm q}-\hat{Q}_\alpha(\tau)\hat{{\cal O}}_\alpha-\chi^+_i \hat{V}_i(\tau)
   \bigr] }\}
   \hat{R}  \overleftarrow{T}\exp\{\;{i} \int_{t_1}^{t}{d\tau\bigl[ \hat{H}_{\rm q}-\hat{Q}_\alpha(\tau)\hat{{\cal O}}_\alpha-\chi^-_i\hat{V}_i \bigr]}\} \end{equation}
The tracing out the environment in the next smaller interval of duration ${\cal T}$ promotes the pseudo-density matrix as 
\begin{equation}
\tilde{\rho}(t+{\cal T}) = \tilde{\rho}(t) +{\cal T}\left(-i [H_{{\rm q}},\tilde{\rho}(t)] - \Gamma[\tilde{\rho}(t)]\right)\end{equation}
where the linear superoperator $\Gamma$ will be evaluated below.
Therefore, the whole FCS kernel can be presented as
\begin{equation}
P(\boldsymbol{\chi}^+,\boldsymbol{\chi}^-) = {\rm Tr}[\tilde{\rho}(t_2)]
\end{equation}
where $\tilde{\rho}(t_2)$ is obtained by solving an evolution equation 
\begin{equation}
\frac{\partial \tilde{\rho}}{\partial t}= -i [H_{{\rm q}},\tilde{\rho}(t)] - \Gamma[\tilde{\rho}(t)]
\end{equation}
with the initial condition $\tilde{\rho}(t_1) = \rho$, $\rho$ being the true density matrix of the system to be measured at the time moment $t_1$. 

Let us evaluate the linear superoperator $\Gamma$. It is contributed by various second-order terms in operators $\hat{V}_i, \hat{Q}_\alpha$. There are contributions proportional to the second, first, and zero power of $\chi^{\pm}$. Let us consider these three contributions separately.

The second order terms involve the correlators of two $\hat{V}_j$ operators.
We denote 
\begin{equation}
\label{eq:notationsnoise}
S^{\pm}_{ij}=\int dt \langle \hat{V}_i(0) \hat{V}_j(t) \Theta(\pm t)\rangle
\end{equation}
and rewrite it as 
\begin{equation}
\Gamma[\rho] = \left(
\chi_i^+\chi_j^+ S_{ij}^- + 
\chi_i^-\chi_j^- S_{ij}^+ 
-\chi_i^-\chi_j^+ (S_{ij}^- +S_{ij}^+)
\right)
\rho
\end{equation}
At this point, it is convenient to  introduce symmetrized noises $S_{ij}$ and the susceptibilities $a_{ij}$,
\begin{eqnarray}
2 S_{ij} &=& S_{ij}^++S_{ij}^-+S_{ji}^-+S_{ji}^+\\
a_{ij} &=& i(S_{ji}^+ - S_{ij}^-)
\end{eqnarray}
With these more physical quantities, we express the sums of correlators 
\begin{eqnarray}
\label{eq:noiserelations}
S_{ij}^\pm+S_{ji}^\pm  = S_{ij} \mp  i\frac{a_{ij}+a_{ji}}{2}\\
S_{ij}^- +S_{ij}^+ =  S_{ij} +i \frac{a_{ij}-a_{ji}}{2}
\end{eqnarray}
to obtain  
\begin{equation}
\Gamma[\rho] = \frac{1}{2}\left(
(\chi_i^+-\chi_i^+)(\chi_j^+-\chi_j^+) S_{ij} + 
i (\chi_i^++\chi_i^+)(\chi_j^+-\chi_j^+) a_{ij}
\right)
\rho
\end{equation}
To make sure that the FCS kernel defines the probability distribution,
we need to require $a_{ij}=0$, no zero-frequency susceptibilities of the detector.
We stress that this is the case of most common electrical measurement. The operators $\hat{V}_i$ in this case are associated with currents or voltages in a dissipative electrical circuit. The zero-frequency susceptibilities in this situation would give current and/or voltage response on vector potential and/or charge passed through a point in a circuit, therefore they are zero by virtue of gauge invariance.

Let us evaluate the first-order contribution. In this case, each term involves a single operator $\hat{{\cal O}}$ and a correlator of $\hat{Q}$, and $\hat{V}$. Adopting the notations (\ref{eq:notationsnoise}), we represent this term as
\begin{equation}
\Gamma[\rho] = \hat{{\cal O}}_\alpha \rho 
\left(\left(S^-_{\alpha i} +S^-_{i\alpha}\right)\chi^+_i - \left(S^+_{i\alpha } +S^-_{i\alpha}\right)\chi^-_i\right)+ 
\rho\hat{{\cal O}}_\alpha \left(\left(S^+_{\alpha i} +S^+_{i\alpha}\right)\chi^-_i - \left(S^+_{\alpha i} +S^-_{\alpha i}\right)\chi^+_i\right)
\end{equation}

Making use of the relations (\ref{eq:noiserelations}), we arrive at
\begin{equation}
\Gamma[\rho] = \hat{{\cal O}}_\alpha \rho \left(\left(S_{\alpha i} +i\frac{a_{ i\alpha}}{2}\right)(\chi^+_i -\chi^-_i) + i\frac{a_{\alpha i}}{2} (\chi^+_i +\chi^-_i)\right)+
\rho\hat{{\cal O}}_\alpha \left(\left(S_{\alpha i} -i\frac{a_{ i\alpha}}{2}\right)(\chi^-_i -\chi^+_i) - i\frac{a_{\alpha i}}{2} (\chi^+_i +\chi^-_i)\right)
\end{equation}

We see that the terms with the sums of counting fields drop and the correctness of FCS approach is guaranteed  provided  $a_{\alpha i}=0$, that is, there are no susceptibilities from the detectors to the measured variables. Again this is the case of a common electrical measurement and is guaranteed by guage invariance. The reverse susceptibilities $a_{i \alpha}$ should be non-zero for the measurement to take place.

The zero-order contribution describes the effect of environment on the dynamics of the measured system. It involves the pairs of the operators $\hat{{\cal O}}$, and reads
\begin{equation}
\Gamma[\rho] = 
S_{\alpha \beta}^-\hat{{\cal O}}_\alpha \hat{{\cal O}}_\beta \rho +
\rho \hat{{\cal O}}_\alpha \hat{{\cal O}}_\beta S_{\alpha \beta}^+
- \hat{{\cal O}}_\alpha \rho \hat{{\cal O}}_\beta (S_{\beta \alpha}^+ + S_{\beta\alpha}^-) 
\end{equation}
It is instructive to separate this expression into two parts. 
The first part is a Lindblad form describing dissipative transitions and decoherence induced by the environment,
\begin{equation}
\Gamma_L[\rho] = 
\left(\frac{1}{2}\left(\hat{{\cal O}}_\beta \hat{{\cal O}}_\alpha \rho +
\rho \hat{{\cal O}}_\beta \hat{{\cal O}}_\alpha \right)
- \hat{{\cal O}}_\alpha \rho \hat{{\cal O}}_\beta\right) C_{\beta \alpha}
\end{equation}
where the hermitian matrix $C_{\beta \alpha}$ is defined as
\begin{equation}
\label{eq:C}
C_{\beta \alpha} =(S_{\beta \alpha}^+ + S_{\beta\alpha}^-) = S_{\beta\alpha}  +i \frac{a_{\beta \alpha}-a_{\alpha \beta}}{2}. 
\end{equation}
One can diagonalize the set of Linblad operators involved. For this, let us present $C$ in the diagonal form, 
\begin{equation} 
C_{\alpha\beta} = \Psi^{*\gamma}_\alpha C_\gamma \Psi^\gamma_{\beta},
\end{equation}
with $\gamma$ labeling its eigenvectors and eigenvalues, and introduce an operator  set 
\begin{equation}
\hat{L}_\gamma = \sqrt{C_\gamma} \Psi^{\gamma}_{\alpha} \hat{{\cal O}}_\alpha.
\end{equation}
In these terms, the contribution reads
\begin{equation}
\Gamma_L[\rho] = 
\frac{1}{2}\left(\hat{L}_\gamma^\dagger  \hat{L}_\gamma\rho +
\rho \hat{L}_\gamma^\dagger  \hat{L}_\gamma\right) 
- \hat{L}_\gamma\rho \hat{L}_\gamma^\dagger 
\end{equation}
The second part gives a renormalization of the system Hamiltonian
by the coupling to the environment. It reads
\begin{equation}
\Gamma_H[\rho] = i [\hat{H}',\rho]; \; \hat{H}'= -\frac{i}{2} (S^-_{\alpha\beta} - S^+_{\alpha\beta}) \hat{{\cal O}}_\alpha \hat{{\cal O}}_\beta
\end{equation}
The  matrix $i(S^-_{\alpha\beta} - S^+_{\alpha\beta})/2$ is Hermitian and in general case cannot be expressed in terms of zero-frequency noises and susceptibilities. With a help of a Kramers-Kronig relation, it can be expressed in terms of those at finite frequency, 
\begin{equation}
\frac{i}{2} (S^-_{\alpha\beta} - S^+_{\alpha\beta}) = \int \frac{d\omega}{2\pi \omega} \left(\frac{a_{\beta\alpha}(-\omega) - a_{\alpha\beta}(\omega)}{2} +i S_{\alpha\beta}(\omega)\right)
\end{equation}
If the environment is in the ground state, this matrix can be reduced to the matrix of the zero-frequency susceptibilities, $i(S^-_{\alpha\beta} - S^+_{\alpha\beta})/2 = a_{ij}$. Since this term can be attributed to the system Hamiltonian, it is not especially interesting for us and we do not discuss it further.

To summarize the results of the derivation of this Section,
the distribution of integrated detector outputs $P(\boldsymbol{s})$ over the time interval $(t_1,t_2)$ is expressed in terms of a pseudo-density matrix $\tilde{\rho}$ that depends on the counting fields $\boldsymbol{\chi}$,
\begin{equation}
P(\boldsymbol{s}) = \int{\frac{d \boldsymbol{\chi}}{2 \pi} \; e^{i\boldsymbol{s}\cdot \boldsymbol{\chi}}\; {\rm Tr}[\tilde{\rho}(\boldsymbol{\chi};t_2)]}. 
\end{equation}
and satisfies the evolution equation
\begin{eqnarray}
\label{eq:FCS}
\frac{\partial \tilde{\rho}}{\partial t}= -i [H_{{\rm q}},\tilde{\rho}(t)] - \frac{1}{2} \chi_i S_{ij} \chi_j \\
-\left(
\hat{{\cal O}}_\alpha \rho \left(S_{\alpha i} +i\frac{a_{ i\alpha}}{2}\right)-
\rho\hat{{\cal O}}_\alpha \left(S_{\alpha a} -i\frac{a_{ i\alpha}}{2}\right)
\right)\chi_i \\
- \left[\frac{1}{2}\left(\hat{{\cal O}}_\beta \hat{{\cal O}}_\alpha \rho +
\rho \hat{{\cal O}}_\beta \hat{{\cal O}}_\alpha \right)
- \hat{{\cal O}}_\alpha \rho \hat{{\cal O}}_\beta\right]\left(S_{\beta \alpha} +i \frac{a_{\beta \alpha}-a_{\alpha \beta}}{2}\right)
\end{eqnarray}
with initial condition $\tilde{\rho}(t_1) = \rho(t_1)$, $\rho(t_1)$ being the density matrix of the system measured.

The noises and susceptibilities involved in this equation are not arbitrary numbers. They should satisfy inequalities that follow from their definition and eventually guarantee that the distribution of the outcomes obtained from the above equation, is positively defined. 

Let us consider а matrix $\check{C}$ (Eq. \ref{eq:C}) with an index $a$ that takes values of detector and operator indices, 
$C_{ab} = S_{ab}+ i(a_{ba}-a_{ab})/2$. All inequalities required are obtained from the condition that the matrix $\check{C}$ is {\it positively defined}, that is, for any vector $\Psi_a$ , $\Psi^*_a C_{ab} \Psi_b >0$

If the vector has a single component, the positivity requires rather obvious inequalities $S_{ii}>0$, $S_{\alpha\alpha}>0$, diagonal noises are positive.
For a two-component vector, in addition to the above conditions, the determinant of the corresponding $2\times2$ matrix must be positive. For two detectors, this restricts cross-noises since the corresponding susceptibilities are 0, $S_{ii}S{jj} > S^2_{ij}$. For detector $i$ and operator $\alpha$, 
this gives the condition 
\begin{equation}
S_{ii}S_{\alpha\alpha} > S^2_{i\alpha} + a^2_{i\alpha}
\end{equation}
that is widely discussed in the context of CWLM\cite{reviewnoise}. Increasingly complex inequalities can be obtained if one considers the vectors with more components\cite{Albert2}.

\section{Drift-diffusion equation}
\label{sec:driftdiffusion}
There is an alternative way to view this equation. 
Let us consider a density matrix in system variables 
and the auxiliary variables $\boldsymbol{s}$ that we have used to represent the integrated detector outputs, $\rho(\boldsymbol{s}_1,\boldsymbol{s}_2)$ where we have made explicit its dependence on the outputs.  
As a matter of fact, the $\boldsymbol{\chi}$-dependent pseudo-density matrix $\tilde{\rho}$ can be regarded as a Fourier-component of this density matrix for coinciding $\boldsymbol{s}_1,\boldsymbol{s}_2$,
\begin{equation}
\tilde{\rho}(\boldsymbol{\chi}) = \int d \boldsymbol{\chi} e^{i \boldsymbol{\chi} \cdot \boldsymbol{s}} \rho(\boldsymbol{s},\boldsymbol{s})
\end{equation}

Performing the inverse Fourier transform, we obtain the following equation
for $\rho(\boldsymbol{s}) \equiv \rho(\boldsymbol{s},\boldsymbol{s})$ (here, $\partial_i \equiv \partial_{s_i}$)

\begin{eqnarray}
\label{eq:drift-diffusion}
\frac{\partial \tilde{\rho}}{\partial t}= -i [H_{{\rm q}},\tilde{\rho}(t)] + \frac{1}{2}  S_{ij} \partial_i \partial_j \rho\\
-i \left(
\hat{{\cal O}}_\alpha \partial_i\rho \left(S_{\alpha i} +i\frac{a_{i\alpha}}{2}\right)-
\partial_i \rho\hat{{\cal O}}_\alpha \left(S_{\alpha a} -i\frac{a_{ i\alpha}}{2}\right)
\right) \\
- \left[\frac{1}{2}\left(\hat{{\cal O}}_\beta \hat{{\cal O}}_\alpha \rho +
\rho \hat{{\cal O}}_\beta \hat{{\cal O}}_\alpha \right)
- \hat{{\cal O}}_\alpha \rho \hat{{\cal O}}_\beta\right]\left(S_{\beta \alpha} +i \frac{a_{\beta \alpha}-a_{\alpha \beta}}{2}\right)
\end{eqnarray}

This equation is of the drift-diffusion type. In the absence of coupling to the quantum system, it describes a Brownian motion in the multi-dimensional space of integrated outputs. In this case, $\rho$ is just a scalar giving the probability of the integrated outcome $\boldsymbol{s}$,
\begin{equation}
P_0(\boldsymbol{s}) = \sqrt{\frac{{\rm det}[S_{ij}] }{2\pi t}}\exp\left(-\frac{s_i s_j (S^{-1})_{ij}}{2t}\right)
\end{equation}
In the presence of coupling, the maximum of this distribution drifts with a velocity that is proportional to the measured values of the operators $\hat{{\cal O}}_\alpha$.

A simple and general solution of the equation (\ref{eq:drift-diffusion}) can be obtained under a rather uninteresting "classical" assumption that all operators $\hat{{\cal O}}_\alpha$ commute with each other.
In this case, the equations for the elements of $\rho$ separate in the basis of the eigenvectors of $\hat{{\cal O}}_\alpha$ . 

The time evolution of a diagonal element $\rho\left(\lbrace{\cal O}_\alpha\rbrace,\boldsymbol{s}\right)$ 
exhibits a simple drift-diffusion behavior,
\begin{equation}
\rho\left(\lbrace{\cal O}_\alpha\rbrace,\boldsymbol{s}\right) = P_0(\boldsymbol{s} - \boldsymbol{v}t),
\end{equation}
with the velocity $v_k \equiv a_{k\alpha} {\cal O}_\alpha$ proportional to the eigenvalues. 

The non-diagonal elements, in addition to drift, are subject to damping due to decoherence and aslo exhibit oscillations due to noise correlations $S_{\alpha k}$, and non-symmetric susceptibilities,

\begin{equation}
\rho\left(\lbrace{\cal O}_\alpha\rbrace,\lbrace{\cal O}'_\alpha\rbrace,\boldsymbol{s}\right) = P_0\left(\boldsymbol{s} - \frac{\boldsymbol{v} +\boldsymbol{v}'}{2} t +i (\boldsymbol{w} - \boldsymbol{w}') t \right) e^{-\Gamma_d t + i \gamma t}.
\end{equation}
Here, $v'_k = a_{k\alpha} {\cal O}'_\alpha$, $w_k = S_{\alpha k}{\cal O}_\alpha$, $w'_k = S_{\alpha k}{\cal O}'_\alpha$, $\Gamma_d = \tfrac{1}{2} S_{\alpha\beta}({\cal O}_\alpha - {\cal O}'_\alpha)({\cal O}_\beta - {\cal O}'_\beta)$, $\gamma = \tfrac{1}{4} a_{\alpha\beta}(({\cal O}_\alpha + {\cal O}'_\alpha) ({\cal O}_\beta - {\cal O}'_\beta) - ({\cal O}_\beta + {\cal O}'_\beta))({\cal O}_\alpha - {\cal O}'_\alpha)$.

The solutions become much more involved in the case of non-commuting $\hat{{\cal O}}_\alpha$.

\section{Lindblad construction derivation}
\label{sec:lindblad}
In this Section, we will 'reverse-engineer' the drift-diffusion equation (\ref{eq:drift-diffusion}) providing its general phenomenological derivation that is mostly based on the positivity of the density matrix utilizing Lindblad construction. This equation is for a density matrix $\hat{\rho}(\boldsymbol{s}_1,\boldsymbol{s}_2)$, where $\boldsymbol{s}$ represents the detector outputs while the rest of the matrix structure is inherited from the measured system. An important additional requirement on the equation is that it does not mix diagonal and non-diagonal components of the matrix this suppressing possible quantum interference of the states with different detector readings.

Let us start with Lindblad construction. Given a set of operators $\hat{A}_i$ and the Hermitian Hamiltonian $\hat{H}$ the positivity of a general density matrix is guaranteed by the following equation (Linblad construction)
\begin{equation}
\label{eq:construction}
\frac{\partial \hat{\rho}}{\partial t} = 
S_{ij} \left( \hat{A}_i \hat{\rho} \hat{A}^\dagger_j  - \frac{1}{2} \hat{A}^\dagger_j\hat{A}_i \hat{\rho} - \frac{1}{2} \hat{\rho} \hat{A}^\dagger_j\hat{A}_i\right) - i \hat{H} \hat{\rho} + i \hat{\rho} \hat{H}
\end{equation}
provided $S_{ij}$ is a {\it positive Hermitian matrix}. At the moment, it is an arbitrary matrix not related to the matrix $S_{ij}$ used in the previous Sections.

Let us specify to the structure 
$\hat{\rho}({\bf s}_1,{\bf s}_2)$. It is convenient to introduce the half-sum and the half-difference of these variables,
\begin{equation}
{\bf s},{\bf d}  \equiv \frac{{\bf s}_1\pm {\bf s}_2}{2}.
\end{equation}
Let us find a Lindblad constuction that does not mix diagonal and non-diagonal matrix elements in $\boldsymbol{s}$. As for the operator set, we choose
$$
\hat{A}_i = \hat{\chi}_i + \hat{B}_i, \; \hat{\chi}_i \equiv i \partial/\partial s_{i} 
$$
This gives three groups of terms.

First group represents the diffusion in the space of detector variables containing the terms quadratic in $\hat{\chi}$.
\begin{equation}
\frac{\partial \hat{\rho}}{\partial t} = 
S_{i j} \left(\hat{\chi}_i \hat{\rho}\hat{\chi}_j - \frac{1}{2} \hat{\chi}_i \hat{\chi}_j \hat{\rho} -  \frac{1}{2} \hat{\chi}_i \hat{\chi}_j \hat{\rho}\right)
\end{equation}
We notice that
\begin{eqnarray*}
<{\bf s}_1|\hat{\rho}\hat{\chi}_i|{\bf s}_2> &=& - i \frac{\partial \hat{\rho}}{\partial s_{2,i}} = -i \frac{1}{2}\left(\frac{\partial}{\partial s_i} - \frac{\partial}{\partial d_i}\right) \hat{\rho} \\
<{\bf s}_1|\hat{\chi}_i\hat{\rho}|{\bf s}_2> &=& i \frac{\partial \hat{\rho}}{\partial s_{1,i}} = i \frac{1}{2}\left(\frac{\partial}{\partial s_i } + \frac{\partial}{\partial d_{i}}\right) \hat{\rho} 
\end{eqnarray*}
With this, the equation for density matrix is represented as 
\begin{equation}
\label{eq:quadric}
\frac{\partial \hat{\rho}}{\partial t} = 
{\rm Re}\left[S_{i j}\right] \frac{1}{2} \frac{\partial}{\partial s_i}\frac{\partial}{\partial s_i}\hat{\rho} +i {\rm Im}\left[S_{i j}\right] \frac{1}{2} \frac{\partial}{\partial d_i}\frac{\partial}{\partial s_j} \hat{\rho}
\end{equation}
We have to require here the absence of the terms with the derivatives with respect to ${\bf d}$. It may seem to require {\it real} and therefore symmetric matrix $S$. 

However, there could be a term in $\hat{H}$ compensating for imaginary part of $S$. This could happen if this part of the Hamiltonian contains two derivative operators, so let us search for it in the most general form
$H =  \bar{C}_{ij} \hat{\chi}_i \hat{\chi}_j$ with a Hermitian $\bar{C}$. This gives the following contribution to the time derivative of the density matrix:
\begin{equation}
\frac{\partial \hat{\rho}}{\partial t} = -i 
\bar{C}_{ij} 
\left(\hat{\chi}_i \hat{\chi}_j \hat{\rho}  -\hat{\rho}\hat{\chi}_i \hat{\chi}_j\right) = -i 
{\rm Re}\left[\bar{C}_{ij}\right] \frac{\partial}{\partial d_i}\frac{\partial}{\partial s_j} \hat{\rho}.
\end{equation}
This is always symmetric with respect to exchange of $i$ and $j$, so it cannot compensate the operator in the second term of Eq. \ref{eq:quadric} which is antisymmetric. Therefore $S$ is indeed a symmetric and real matrix.

Second group of terms mixes $\hat{\chi}$ and $\hat{B}$.
\begin{equation}
\frac{\partial \hat{\rho}}{\partial t} = 
S_{i j} \left( \hat{\chi}_i \hat{\rho} \hat{B}^{\dagger}_j + \hat{B}_i \hat{\rho} \hat{\chi}_j - \frac{1}{2}\left(\hat{\chi}_i \hat{B}_j + \hat{B}^\dagger_i \hat{\chi}_j \right)\hat{\rho} -\frac{1}{2} \hat{\rho}\left(\hat{\chi}_i \hat{B}_j + \hat{B}^\dagger_i \hat{\chi}_j \right) \right)
\end{equation} 
We collect the terms proportional to the derivatives of ${\bf s}$
$$
\frac{i}{4} 
S_{i j} \left( \frac{\partial \hat{\rho}}{\partial s_i} \left(3 \hat{B}^\dagger_j + \hat{B}_j\right) - \left(3 \hat{B}_j + \hat{B}^\dagger_j\right) \frac{\partial \hat{\rho}}{\partial s_i}\right)
$$
and to the derivatives of ${\bf d}$:
$$
\frac{i}{4} 
S_{i j} \left(\frac{\partial \hat{\rho}}{\partial d_i} \left(\hat{B}^\dagger_j -\hat{B}_j \right) - \left(\hat{B}^\dagger_j -\hat{B}_j \right) \frac{\partial \hat{\rho}}{\partial d_i}\right).
$$

We do not like the terms proportional to the derivatives of ${\bf d}$. Let us try to compensate those by a proper choice of an addition to the Hamiltonian. We seek for it in the form $\hat{H} = - \sum_{i} \hat{\chi}_i \hat{D}_i$,
$\hat{D}_i$ being Hermitian operators. Its contribution to the time derivative of the density matrix reads
\begin{equation}
\frac{\partial \hat{\rho}}{\partial t}  = -\frac{1}{4} \left( \hat{D}_i \frac{\partial \hat{\rho}}{\partial s_i} +  \frac{\partial \hat{\rho}}{\partial s_i}\hat{D}_i\right) - \frac{1}{4} \left( \hat{D}_i \frac{\partial \hat{\rho}}{\partial d_i} - \frac{\partial \hat{\rho}}{\partial d_i} \hat{D}_i \right)
\end{equation}
To cancel the terms we dislike we need to set 
$$
\hat{D}_i = i 
S_{ij} 
\left( \hat{B}_j -\hat{B}^\dagger_j \right)
$$
Summing up the terms from the Linblad form and the Hamiltonian, we obtain 
\begin{equation}
\frac{\partial \hat{\rho}}{\partial t} = i\left(\frac{\partial \hat{\rho}}{\partial s_i} \hat{K}^\dagger_i - \hat{K}_i\frac{\partial \hat{\rho}}{\partial s_i} \right);\; \hat{K}_i = 
S_{ij} \hat{B}_j
\end{equation} 
We can also separate  $\hat{K}_i$ into two Hermitian operators, $\hat{K}_i = (\hat{R}_i - i \hat{D}_i)/2$, with this
\begin{equation}
\frac{\partial \hat{\rho}}{\partial t} = \frac{i}{2}\left[\frac{\partial \hat{\rho}}{\partial s_i}, \hat{R}\right] 
+ \frac{1}{2} \left[\frac{\partial \hat{\rho}}{\partial s_i}, \hat{D}\right]_+
\end{equation} 
where, as we will see soon, the first term is associated with the effect of cross-noises between the detector variables and the fields acting on the measured system, while the second term is associated with the susceptibilities.

The third group of terms represents the effect of the measurement on the decoherence and relaxation of the quantum system.
\begin{equation}
\frac{\partial \hat{\rho}}{\partial t}  = S_{ij} \left( \hat{B}_i \hat{\rho} \hat{B}^\dagger_j  - \frac{1}{2} \hat{B}^\dagger_j\hat{B}_i \hat{\rho} - \frac{1}{2} \hat{\rho}\hat{B}^\dagger_j\hat{A}_i\right) =  \hat{B}_i \hat{\rho} \hat{K}^\dagger_i -\frac{1}{2} \hat{B}_i \hat{K}^\dagger_i \hat{\rho}-\frac{1}{2}  \hat{\rho}\hat{B}_i \hat{K}^\dagger_i
\end{equation}

We can bring everything together to a relatively compact form:
   
\begin{equation}
\label{eq:leq}
\frac{\partial \hat{\rho}}{\partial t} = 
\frac{1}{2} S_{ij} \partial_i \partial_j \hat{\rho} 
+ i \left(\partial_i \hat{\rho} \hat{K}^\dagger_i- \hat{K}_i \partial_i \hat{\rho}\right) 
+ \frac{1}{2} \left[\hat{B}_i,\hat{\rho} \hat{K}^\dagger_i\right] 
+ \frac{1}{2} \left[\hat{B}_i \hat{\rho}, \hat{K}^\dagger_i\right]
\end{equation}

Let us now compare this with Eq. \ref{eq:drift-diffusion} term by term. The comparison of the first group of terms shows that the matrix $S$ is nothing but the noise matrix of the detectors. The comparison of the second group gives
\begin{equation}
\hat{D}_i = a_{i\alpha} \hat{{\cal O}}_\alpha,\; \hat{R}_i = 2 S_{i\alpha} \hat{{\cal O}}_\alpha,\; \hat{K}_i = \left(S_{\alpha i} -i \frac{a_{i\alpha}}{2}\right )
\end{equation}
so the operators $\hat{R},\hat{D}$ are indeed associated with the cross-noises and susceptibilities, respectively.

The third group of terms in Eq. \ref{eq:leq} gives the minimum decoherence and dephasing that is associated with the measurement, or, in other words, to the input noises acting on the detector and corresponding susceptibilities. The contributions can also come from other sources that are not related to the measurement. They can be added to the Lindblad constuction (\ref{eq:construction}) as a set of operators ${\cal O}_\alpha$ with a {\it positively} defined Hermitian matrix. With this, we obtain an important result 
\begin{equation}
S_{\alpha \beta } +i \frac{a_{\beta \alpha}-a_{\alpha \beta}}{2} > \left(S_{\alpha i} - i \frac{a_{i\alpha}}{2}\right) (S^{-1})_{ij} \left(S_{j \beta} - i \frac{a_{i\beta}}{2}\right)
\end{equation}
Here, the inequality sign implies that the difference of the matrices on both sides is a positively defined matrix, and the right hand side represents the minimum contribution to the decoherence/dephasing. 

Naturaly, the same inequality may be derived from the positivity of the matrix $\check{C}$ discussed in the previous sections. 

\section{Output rescaling and separation}
\label{sec:separation}
Till this moment, we assume general linear detection working with an arbitrary noise matrix $S_{ij}$. Since the detection is linear, we can redefine the detector outputs taking arbitrary combinations of those. An orthogonal transformation of the outputs brings the noise matrix to the diagonal form. This separates the detectors, their noises are now independent. The rescaling of the separated outputs brings the diagonal noises to the same value $S$. It is possible to set $S=1$. However, this implies the rescaling of the outputs in such a way that all of them have dimension $sec^{1/2}$. We find this rather inconventient so we prefer to work with a dimensionful $S$.

Such redefinition of the outputs simplifies the equations to some extent. The resulting equations are obtained by substitution $S_{ij} = S \delta_{ij}$. In particular, Eq. \ref{eq:leq} takes the form

\begin{equation}
\label{eq:singleLinblad}
S^{-1} \frac{\partial \hat{\rho}}{\partial t} =  \frac{1}{2} \partial_i\partial_i \hat{\rho} + i (\partial \hat{\rho} \hat{B}_i^\dagger - \hat{B}_i \partial \hat{\rho}) + \frac{1}{2} \left[\hat{B}_i,\hat{\rho} \hat{B}_i^\dagger\right] 
+ \frac{1}{2} \left[\hat{B}_i \hat{\rho}, \hat{B}^\dagger_i\right]
\end{equation}

\section{Discrete update}
\label{sec:discreteupdate}
In this Section, we will look at the resulting equations from a different point of view: we will introduce a discrete process, a step-by-step update of the density matrix of the system and detector outputs. As we will see in the next Section, this update can be made stochastic giving stochastic trajectories of in the space of integrated detector outputs. The actual $\hat{\rho}({\bf s})$ is then obtained by averaging over different realizations of trajectories. One motivation for considering the stochastic update is that it can be an efficient numerical strategy to solve the drift-diffusion equation. An alternative strategy would involve a discretization of the output space and solving at the resulting multi-dimensional mesh with a lot of nodes. Another motivation is that the stochastic update process can be made to mimic the time-line of an actual experimental run where random outputs of the detectors are quasi-continuously measured.

The stochastic update was considered in \cite{trajectory,CWLM1}.
Here, we present its generalization to general situation of multi-detector measurement of an arbitrary quantum system. 

We start by noting that an update that reproduces the drift-diffusion equation can be organized in a variety of ways. We chose a physical but rather general way. We separate the detectors as in the previous section and concentrate on a single detector. We introduce an auxiliary quantum system for this particular detector. At each update step, we first prepare the auxiliary system in an initial state characterized by a certain density matrix $\hat{R}$. Then we switch on an interaction between 
the auxiliary system, the system to be measured, and the detector variable $\chi$ and let the unitary evolution to take place during a time interval $dt$. 

The idea is to keep $dt$ small so that the change of the system density matrix is small $\propto dt$, and to choose the form of interaction in such a way as to reproduce the contribution of this particular detector into Eq. \ref{eq:singleLinblad}. Generalization to many detectors is straightforward: since the contributions of the detectors add, at each update step we run the procedure described for     
all auxilliary systems representing the detectors. The resulting update does not depend on the order of the procedures with an accuracy $\propto (dt)^2$. The sources of decoherence and relaxation not related to the measurement may be incorporated in a similar way with using the auxiliary systems where there is no interaction with the detector variable.

To understand the requirements on the interaction and to make convenient choices is thus enough to concentrate on a single Linblad operator
$\hat{\chi} + \hat{B}$. It is convenient to organize the update in such a way that the interaction with $\hat{B}$ comes first and them the interaction with the detector variable $\hat{\chi}$ takes place. The update of the density matrix is then defined as follows:
\begin{equation} 
\hat{\rho}_{new} = {\rm Tr}_a\left[ \hat{U} \hat{\rho}_{old} \hat{R} \hat{U}^{-1}\right]
\label{eq:update}
\end{equation}
 with the unitary evolution operator $\hat{U}=\exp\lbrace -i \hat{V}_\chi \rbrace \exp\lbrace -i \hat{V}_B \rbrace$ and 
 the interaction $\hat{V}_{\chi,B}$ assuming the following form
 \begin{equation}
\hat{V}_\chi= (S dt)^{1/2} \hat{\chi} \hat{c} \; ; \hat{V}_B = (S dt)^{1/2}\left( \hat{B} \hat{b}^{\dagger} + \hat{B}^\dagger \hat{b} \right) 
 \end{equation}
and the trace is over the dergrees of freedom of the auxiliary system. At the moment, Hermitian $\hat{c}$ and generally non-Hermitian $\hat{b}$ are abritrary operators in the space of the auxiliary system, with only condition of their zero expectation values $\langle \hat{c} \rangle =0, \langle\hat{b}\rangle=0$ (Here, 
$\langle \hat{A} \rangle \equiv {\rm Tr}_a\left[\hat{A} \hat{R}\right]$. To derive the evolution equation, we need to expand $\hat{U}$ up to the second order in $(S dt)^{1/2}$.
With this, we obtain
\begin{eqnarray*}
S^{-1} \frac{\partial \hat{\rho}}{\partial t} = \\
-\frac{1}{2}  \langle\hat{c}^2
\rangle  \left[\hat{\chi},\left[ \hat{\chi},\hat{\rho}\right]\right] +\\
- \frac{1}{2} \langle\hat{b}^2
\rangle  \left[\hat{B}^\dagger,\left[ \hat{B}^\dagger,\hat{\rho} \right]\right] 
- \frac{1}{2} \langle \hat{b}^{\dagger 2}
\rangle  \left[\hat{B},\left[ \hat{B},\hat{\rho} \right]\right] +\\
+ \langle\hat{b}{\hat{b}^\dagger}
\rangle \left( -\frac{1}{2} \hat{B}^\dagger \hat{B} \hat{\rho}  -\frac{1}{2} \hat{B}^\dagger \hat{B} \hat{\rho} + \hat{B} \rho \hat{B}^\dagger \right) 
+ \langle\hat{b}^\dagger{\hat{b}}
\rangle \left( -\frac{1}{2} \hat{B} \hat{B}^\dagger \hat{\rho}  -\frac{1}{2} \hat{B} \hat{B}^\dagger \hat{\rho} + \hat{B}^\dagger \rho \hat{B} \right) \\
 - \langle\hat{c}\hat{b}
\rangle \hat{B}^\dagger \left[\hat{\chi},\hat{\rho}\right] + \langle\hat{b}\hat{c}
\rangle \left[\hat{\chi},\hat{\rho}\right] \hat{B}^\dagger 
 - \langle\hat{c}\hat{b}^\dagger
\rangle \hat{B} \left[\hat{\chi},\hat{\rho}\right] + \langle\hat{b}^\dagger\hat{c}
\rangle \left[\hat{\chi},\hat{\rho}\right] \hat{B}.
\end{eqnarray*}
Comparing this with Eq. \ref{eq:singleLinblad},
we recognize we have to require 
\begin{equation}
\label{eq:relations}
\langle\hat{c}^2
\rangle=\langle\hat{b}{\hat{b}^\dagger}
\rangle=\langle\hat{b}\hat{c}
\rangle=\langle\hat{c}\hat{b}^\dagger
\rangle=1;\; \langle\hat{b}^2
\rangle=\langle{\hat{b}^{\dagger 2}}
\rangle=\langle\hat{c}\hat{b}
\rangle=\langle\hat{b}^\dagger\hat{c}
\rangle=\langle\hat{b}^\dagger\hat{b}
\rangle=0.
\end{equation}

Those are the only conditions on the corresponding operators, otherwise they can be chosen in an arbitrary way.
We will specify two simple choices below. Yet before this let us present a greater simplification of the method under description. In fact, it is not necessary to deterministically update the whole density matrix that involves the measured system and the detector variables. Equivalently, one can update the system density matrix only while producing at each step a stochastic detector output.

\section{Stochastic trajectories}
\label{sec:trajectories}
To see this possibility, let us rewrite Eq. \ref{eq:update} in the form that explicates eigenstates of the operator $\hat{c}$,

\begin{eqnarray}
\rho_{new} = \sum_c  \exp\lbrace - i (S dt)^{1/2} \hat{\chi} \hat{c} \rbrace  {\cal L}_c \rho_{old}  \exp\lbrace - i (S dt)^{1/2} \hat{\chi} \hat{c} \rbrace; \\
 {\cal L}_c \rho_{old}  \equiv \langle c| \exp\lbrace -i (S dt)\left( \hat{B} \hat{b}^{\dagger} + \hat{B}^\dagger \hat{b} \hat{\rho}_{old}\right\rbrace \hat{R} \exp\lbrace i (S dt)\left( \hat{B} \hat{b}^{\dagger} + \hat{B}^\dagger \hat{b}\right)\rbrace|c
\rangle
\end{eqnarray}
If we write in the density matrix the detector variables explicitly $\hat{\rho} \to \hat{\rho}(s,s')$, and concentrate on diagonal elements, $s=s'$, we see that in the course of the update the $s$ coordinate of any such element  
is shifted by a value proportional to an eigenvalue of $\hat{c}$,
\begin{equation}
\hat{\rho}_{new}(x,x) = \sum_c {\cal L}_s \rho_{old}(x-(Sdt)^{1/2} c,x - (Sdt)^{1/2}c).
\end{equation}
This gives us an idea to regard $c$ as a {\it random variable}. At each update step this variable is generated from the distribution $P(c) = {\rm Tr}\left[({\cal L}_c \hat{r}) \right] $ (the trace here is over the system variables)
and contributes to the time-dependent intergated output $s(t)$. The successive updates thus form a {\it stochastic trajectory} in the space of the outputs, $s(t)$.
So we do not have to worry about $s$-dependence of the density matrix any more since this is certain for a certain trajectory. Instead, we can work with a {\it stochastic} density matrix $\hat{r}$ in the system variables that gets an $c$-dependent update. The actual density matrix 
$\hat{\rho}(s,s;t)$ is obtained by averaging over all stochastic trajectories that end in the point $s$.  
To summarize, the update equations become 
\begin{eqnarray}
\label{eq:fullupdate}
s_{new} &=& s_{old} + (S dt)^{1/2} c; \\
c {\rm \ \ is \ \ random \ \ with \ the \ distribution: \ \ } P(c) = {\rm Tr}\left[({\cal L}_c \hat{r}) \right]  \\
\hat{r}_{new} &=& \frac{{\cal L}_c \hat{r}_{old}}
{P(c)}
\end{eqnarray}

We remind that for N detectors one has to repeat the update 
for each detector at each time interval $dt$, promoting ${\bf s}$ with random ${\bf c}$ in N directions. As mentioned, for the terms of the relevant order $(Sdt)$ the order of these updates does not matter.

In the following two subsections, we describe two concrete examples of the auxiliary systems and corresponding updates.

\subsection{Oscillator update} 
In this case, the possible states of the auxilliary system are those of a harmonic oscillator and the  operators $\hat{b}$, $\hat{b}^\dagger$ are conventional annihilation/creation operators of the oscillator. Initially, the oscillator is prepared in the vacuum state, $\hat{R}=|0\rangle\langle 0|$. 
The operator $\hat{c}$ can be associated with the oscillator coordinate, $\hat{c} = \hat{b} +\hat{b}^\dagger$. This choice satisfies the relations (\ref{eq:relations}).

Conveniently, the distribution of $c$ is closer to Gaussian in the limit $dt \to 0$,
\begin{equation}
P(c) \approx |\langle c|0\rangle|^2 = G(c) \equiv (2\pi)^{-1/2} exp(-c^2/2)
\end{equation}

It is constructive to specify the full update equation analytically for two cases: no cross-noise, $\hat{B}=-\hat{B}^+ = - \hat{D}/2S$, and no succeptibility, $\hat{B}=\hat{B}^+ = \hat{R}/2S$

For no cross-noise limit, the natural basis is that of eigenfunctions of $\hat{D}$, that we label with $a,b,..$. The unitary part of the update shifts the wavefunction of the oscillator in coordinate space by values proportional to $D_a$. The distribution of $c$ is a composition of shifted Gaussians with weights equal to probabilities to find the system in state $a$
\begin{equation} 
P(c) = r_{aa} G(c-(dt/S)^{1/2} D_a). 
\end{equation}
The density matrix update involves the shifts corresponding both indices,
\begin{equation}
r^{ab}_{new} = r^{ab}_{old} \sqrt{G(c-(dt/S)^{1/2} D_a)G(c-(dt/S)^{1/2} D_b)}/P(c)
\end{equation}

For no-susceptibility limit, the relevant basis is of the eigenfunctions of $\hat{R}$. The unitary part of the update shifts the wave function of the oscillator in momentum space. This does not modify the $P(c)$.
The whole update is unitary 
\begin{equation}
r^{ab}_{new} = r^{ab}_{old} \exp(i c (dt/S)^{1/2} (R_a - R_b))
\end{equation}
yet stochastic owing to the randomness of $c$. If one knows $s(t)$, the measurement can be "undone" in this situation\cite{CWLM26}.

\subsection{Qubit update}
The simplest auxiliary system is a qubit encompassing two quantum states. The relations (\ref{eq:relations}) are satisfied if $\hat{c} = \hat{\sigma}_x$, $\hat{b} = (\sigma_x + i \sigma_y)$ and the initial state is polarized in z-direction ($\sigma$-matrices are in the space of the qubit). Two possible random outcomes are therefore $c=\pm 1$.

Let us explicate the update analytically in two limits. For the no cross-noise limit, the unitary part of the update rotates the qubit spin about the y-axis with the angles proportional to the eigenvalues of $\hat{D}$ The probabilities of $c=\pm 1$ outcome read
\begin{equation}
P(c)=\frac{1}{2}\left(1 - \sin((dt/S)^{1/2}D_a)r^{aa}_{old}\right)
\end{equation}
and the whole update is expressed as
\begin{equation}
r^{ab}_{new} = r^{ab}_{old} (\cos((dt/S)^{1/2}(D_a-D_b)/2) - c \cos((dt/S)^{1/2}(D_a+D_b)/2)/(2 P(c))
\end{equation}
In the no susceptibility limit, the probabilities of both outcomes are equal, and the whole update is random and unitary,
\begin{equation}
r^{ab}_{new} = r^{ab}_{old} \exp(i c (dt/S)^{1/2} (R_a - R_b))
\end{equation} 
The qubit update can be expressed analytically in terms of operators $\hat{B}, \hat{B}^+$ only, yet the expression is to cumbersome to be instructive.

\section{Conclusions}
\label{sec:conclusions}
In conclusion, we have established a general framework for the description of a CWLM of an arbitrary quantum system by an arbitrary number of the detectors. 
We have compared different approaches to the problem and demostrated their equivalence. The approaches include the full counting statistics (FCS) evolution equation a for pseudo-density matrix (Eq. \ref{eq:FCS}), the drift-diffusion equation (Eqs. \ref{eq:drift-diffusion} , \ref{eq:leq}) for a density matrix in the space of integrated outputs, and discrete stochastic updates (Eq. \ref{eq:fullupdate}). We provide the derivation of the underlying equations from microscopic approach based on full counting statistics method (Section \ref{sec:FCS}), a phenomenological approach based on Lindblad construction (Section \ref{sec:lindblad}), and interaction with auxiliary quantum systems representing the detectors (Sections \ref{sec:discreteupdate},\ref{sec:trajectories}). We give the necessary conditions on the phenomenological susceptibilities and noises that guarantee the unambiguous interpretation of the measurement results and the positivity of density matrix. 

The applicability of the framework is restricted by a Markov assumption: no delay of susceptibilities and no time correlation of noises at the time scale of quantum dynamics. Different methods are required to treat the effects of delay and time correlations at quantum level.  
However, the framework can be easily extended  to incorporate delays at classical level. It can be also extended  to describe various quantum feedback schemes where the quantum system is subject to manipulation, and the  decision on the way to manipulate is based on the values of detector outputs. This will be addressed in future work.
\bibliography{LB}

\end{document}